\documentclass[twocolumn,secnumarabic,showpacs,amssymb, amsmath,tightenlines,aps,prl]{revtex4}
\usepackage{longtable}
\usepackage{bm}
\usepackage{epsfig}
\usepackage{graphicx}

\begin{document}

\title[Importance of In-Plane Anisotropy in the
Quasi 2D BaNi$_{2}$V$_{2}$O$_{8}$] {Importance of In-Plane
Anisotropy\\ in the Quasi Two-Dimensional Antiferromagnet
BaNi$_{2}$V$_{2}$O$_{8}$}
\author{W. Knafo$^{1,2}$, C. Meingast$^{1}$, K. Grube$^{1}$,
S. Drobnik$^{1,2}$, P. Popovich$^{1,2}$, P. Schweiss$^{1}$, P.
Adelmann$^{1}$, Th. Wolf$^{1}$, and H. v. L\"{o}hneysen$^{1,2}$}

\address{$^{1}$ Forschungszentrum Karlsruhe, Institut f\"{u}r Festk\"{o}rperphysik, D-76021 Karlsruhe, Germany\\
$^{2}$ Physikalisches Institut, Universit\"{a}t Karlsruhe, D-76128
Karlsruhe, Germany}

\date{\today}

\begin{abstract}

The phase diagram of the quasi two-dimensional antiferromagnet
BaNi$_{2}$V$_{2}$O$_{8}$ is studied by specific heat, thermal
expansion, magnetostriction, and magnetization for magnetic fields
applied perpendicular to $\mathbf{c}$. At $\mu_0H^{*}\simeq1.5$ T,
a crossover to a high-field state, where $T_N(H)$ increases
linearly, arises from a competition of intrinsic and field-induced
in-plane anisotropies. The pressure dependences of $T_N$ and
$H^{*}$ are interpreted using the picture of a pressure-induced
in-plane anisotropy. Even at zero field and ambient pressure,
in-plane anisotropy cannot be neglected, which implies deviations
from pure Berezinskii-Kosterlitz-Thouless behavior.

\end{abstract}


\maketitle

The study of quasi two-dimensional (2D) magnetic systems
\cite{dejongh90} continues to be a focus of theoretical and
experimental investigations, motivated in large part by the
discovery of high-temperature superconductivity in the quasi 2D
cuprates. Further, the search for a magnetic system exhibiting
true Berezinskii-Kosterlitz-Thouless (BKT) behavior, initially
proposed for 2D XY magnetic systems
\cite{berezinskii71kosterlitz73}, has been elusive and has only
been seen in superfluid and superconducting films
\cite{minnhagen87}. Theoretical studies indicate that BKT behavior
can also be expected for 2D Heisenberg systems with a small
easy-plane XY anisotropy \cite{cuccoli03}. Two recent experimental
papers suggest that BaNi$_{2}$V$_{2}$O$_{8}$ may in fact be a
physical realization of such a system \cite{rogado02,heinrich03}.
BaNi$_{2}$V$_{2}$O$_{8}$ has a rhombohedral structure (space group
R$\overline{3}$) and its magnetic properties arise from a
honeycomb-layered arrangement of spins $S=1$ at the Ni$^{2+}$
sites. The quasi 2D properties are due to a strong
antiferromagnetic Heisenberg superexchange $J$ in the NiO
honeycomb layers. Long range antiferromagnetic ordering, which
would be precluded in a purely 2D Heisenberg system, sets in below
the N\'{e}el temperature $T_N\simeq50$ K \cite{rogado02} because
of small additional energy scales, which we include in the
following Hamiltonian:
\begin{eqnarray}
&\mathcal{H}&=-\sum_{i,j}J\mathbf{S}_i\cdot\mathbf{S}_j-\sum_{i,j'}J'\mathbf{S}_i\cdot\mathbf{S}_{j'}\nonumber\\
&&+\sum_{i}D_{XY}(S_i^z)^2-\sum_{i}D_{IP}(S_i^{a})^2-\sum_{i}\mu_0\mathbf{H}\cdot\mathbf{S}_i.\:\:\:\:\:\:\:
\label{hamiltonian}
\end{eqnarray} The planar XY anisotropy $D_{XY}\simeq1$ meV is a
factor of 10 smaller than $J$ \cite{reznik06} and confines the
spins to lie within the honeycomb layers (easy plane). If this
were the only additional term in Eq. (\ref{hamiltonian}), a true
BKT transition could be expected within each 2D layer
\cite{cuccoli03}. However, a real crystal is always
three-dimensional (3D) and a very small interlayer exchange $J'$
ultimately leads to a crossover from 2D to 3D correlations, and
then to a 3D ordering transition \cite{cuccoli03}. The value of
$J'$ is unknown for the present case \cite{reznik06}; however, the
extremely small signal at $T_N$ in the specific heat
\cite{rogado02} suggests that $J'/J$ is very small (typically,
$J'/J$ is in the range 10$^{-2}$-10$^{-6}$ in quasi 2D systems
\cite{dejongh90}). The in-plane anisotropy $D_{IP}$ is estimated
by 4*10$^{-3}$ meV \cite{reznik06} and acts to align the spins
along one of the three equivalent hexagonal easy $a$-axes
\cite{rogado02}. The last term in Eq. (\ref{hamiltonian}) includes
the effect of a magnetic field $H$, which also acts as an
effective anisotropy \cite{dejongh90,knafo07}.

In this Letter, we study the ($T,H$) phase diagram of
BaNi$_{2}$V$_{2}$O$_{8}$ for magnetic fields applied within the
honeycomb planes. The combination of specific heat, thermal
expansion, magnetostriction, and magnetization allows us to
calculate the uniaxial pressure dependences of $T_N$, $J$, and of
a crossover field, $H^*$, related to spin alignment. In
particular, we find a strong anisotropy of the pressure
dependences of $T_N$ in the easy plane, which is related to the
spin direction. This directly underlines the importance of the
in-plane anisotropy term $D_{IP}$ in Eq. (\ref{hamiltonian}) for
establishing 3D long range order in BaNi$_{2}$V$_{2}$O$_{8}$. Our
analysis and comparison with literature indicates that the
obtained ($T,H$) phase diagram may be generic to quasi 2D
antiferromagnets with a degenerate number of easy axes and with
$J\gg D_{XY}\gg D_{IP}, J'$. It would be useful if the theoretical
models for quasi-2D Heisenberg systems, which already consider XY
anisotropy and inter-plane coupling \cite{cuccoli03}, would also
incorporate this in-plane anisotropy.

Single crystals of BaNi$_{2}$V$_{2}$O$_{8}$ (42.8 mg) and
BaNi$_{0.8}$Mg$_{1.2}$V$_{2}$O$_{8}$ (4.85 mg) were grown in
fluxes composed of BaCO$_3$, NiO, MgO, and V$_2$O$_5$, using
Al$_2$0$_3$ crucibles. Specific heat was measured using a PPMS
from Quantum Design \cite{lashley03}. Thermal expansion and
magnetostriction were measured using a high-resolution capacitive
dilatometer with temperature and field sweep rates of 20 mK/s and
0.5 T/min, respectively. The cell is rotatable, allowing both
longitudinal and transverse measurements. Magnetization was
measured using a MPMS from Quantum Design. For all measurements,
the magnetic field was applied parallel to the easy plane.

\begin{figure}[b]
    \centering
    \epsfig{file=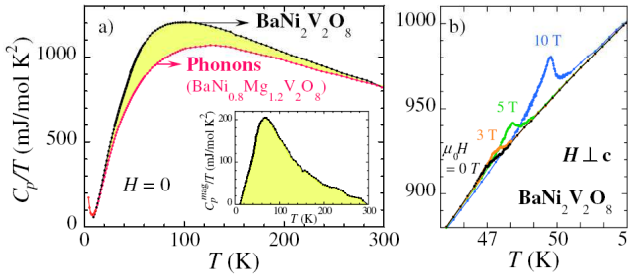,width=83mm}
    \caption{(color online) (a) Specific heat of BaNi$_{2}$V$_{2}$O$_{8}$ and
    BaNi$_{0.8}$Mg$_{1.2}$V$_{2}$O$_{8}$, plotted as $C_p/T$ versus $T$;
    the estimated magnetic specific heat of BaNi$_{2}$V$_{2}$O$_{8}$
    is plotted in the insert. (b) $C_p/T$ versus $T$ of
    BaNi$_{2}$V$_{2}$O$_{8}$ with $H\perp c$.}
    \label{spec_heat}
\end{figure}

\begin{figure}[t]
    \centering
    \epsfig{file=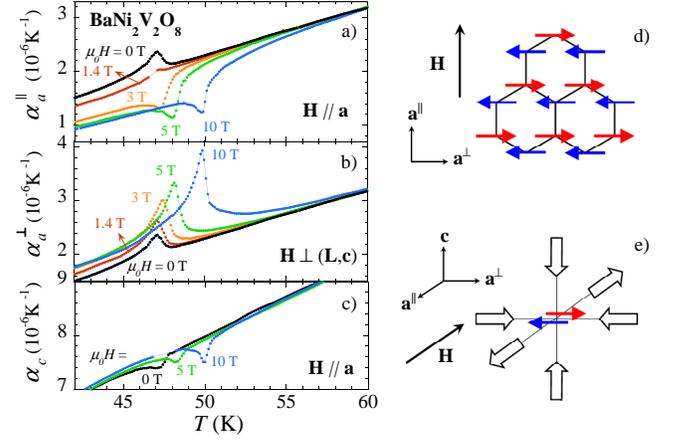,width=58mm,angle=270}
    \caption{(color online) (a-c) $a-$ and $c-$axes thermal expansivity of BaNi$_{2}$V$_{2}$O$_{8}$ versus temperature,
    the field being applied parallel to the easy-plane, i.e. $H\perp
    c$. (d-e) Schematics of the magnetic structure of the Ni$^{2+}$
    ions in the plane and of the field-induced lattice distortion (below $T_N$).}
    \label{th_exp}
\end{figure}

Fig. \ref{spec_heat} (a) shows the 0-T specific heat $C_p$ of
BaNi$_{2}$V$_{2}$O$_{8}$ in a $C_p/T$ versus $T$ plot. An
estimation of the phonon contribution $C_p^{ph}$ is made using the
specific heat of BaNi$_{0.8}$Mg$_{1.2}$V$_{2}$O$_{8}$
\cite{notephonons}. The resulting magnetic specific heat
$C_p^{mag}=C_p-C_p^{ph}$ is shown in the insert of Fig.
\ref{spec_heat} (a), where $C_p^{mag}/T$ is plotted as a function
of $T$. The integrated entropy $\Delta S_{mag}\simeq24$ J/mol K is
roughly equal to the value $2R$ln$3\simeq18$ J/mol K expected for
the $S=1$ Ni$^{2+}$ ions. In this plot, there is little signature
of the transition at $T_N$; rather, $C_p^{mag}/T$ has a broad
maximum at $T_{max}\simeq70$ K which is attributed to the build-up
of 2D correlations and corresponds to the highest magnetic energy
scale $J$. Fig. \ref{spec_heat} (b) shows a blow-up of the $C_p$
anomaly related to the ordering at $T_N$. At $H=0$, we clearly
observe a tiny peak yielding $T_N=47.4\pm0.1$ K, defined as the
locus of the minimum of $\partial (C/T)/\partial T$
\cite{noterogado}. The magnetic entropy $\Delta S_N\simeq13$
mJ/mol K gained at $T_N$ is only a tiny fraction of the total
magnetic entropy, the very small ratio $\Delta S_N/\Delta S_{mag}
\simeq 7*10^{-4}$ \cite{noterogado} being a consequence of the
strong 2D Heisenberg character of the system \cite{bloembergen77}.
Application of a magnetic field in the easy plane leads, at 10 T,
to the increases of $T_N$ by about 3 K and of $\Delta S_N$ by a
factor 4.

The linear thermal expansivity $\alpha=(1/L)\partial L/\partial T$
is shown in Fig. \ref{th_exp} (a-c) for $T$ close to $T_N$ and
$0\leq \mu_0H\leq10$ T. Data for three configurations are
presented: a longitudinal one, $\alpha_a^\parallel$ (
$\mathbf{L}\parallel \mathbf{H}\parallel \mathbf{a}$), and two
transverse ones, $\alpha_a^\perp$ ($\mathbf{L}\parallel
\mathbf{a}$ and $\mathbf{H}\perp (\mathbf{L},\mathbf{c})$), and
$\alpha_c$ ($\mathbf{L}\parallel \mathbf{c}$ and
$\mathbf{H}\parallel \mathbf{a}$). At $H=0$, $T_N$ is
characterized by the jumps $\Delta \alpha_a>0$ (Fig. \ref{th_exp}
(a-b)) and $\Delta \alpha_c<0$ (Fig. \ref{th_exp} (c)).
Application of a magnetic field in the easy plane induces an
increase of $T_N$, defined at the extremum of $\partial
\alpha/\partial T$, in agreement with the specific heat data. The
field induces a sign change of $\Delta \alpha$ for the
longitudinal configuration (Fig. \ref{th_exp} (a)), while the sign
of $\Delta \alpha$ does not change for the transverse
configurations (Fig. \ref{th_exp} (b-c)).

\begin{figure}[t]
    \centering
    \epsfig{file=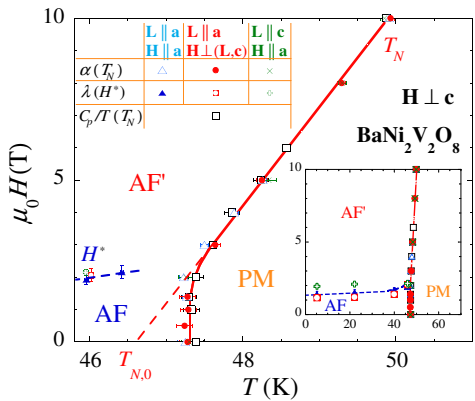,width=62mm}
    \caption{(color online) Phase diagram obtained with $\mathbf{H}\perp\mathbf{c}$. PM, AF, and AF' denote the paramagnetic, the low- and the high-field
    antiferromagnetic phases, respectively.}
    \label{phase_diag}
\end{figure}

\begin{figure}[b]
    \centering
    \epsfig{file=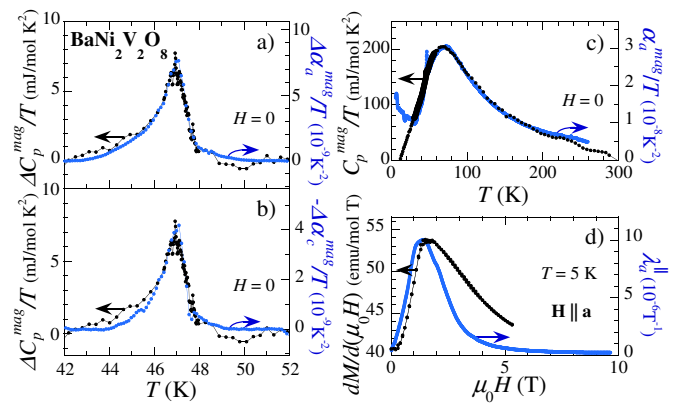,width=87mm}
    \caption{(color online) (a-c) Scalings of specific heat and thermal expansivity (with a minus sign for the $c$-axis) used to determine
    the zero-field uniaxial pressure dependences of $T_N$ and $J$.
    (d) Scaling of $\partial M/\partial (\mu_0H)$ and $\lambda_a^\parallel$ versus $H$ at
    5 K.}
    \label{ehrenfest}
\end{figure}

Fig. \ref{phase_diag} shows the values of $T_N$ extracted from
specific heat and thermal expansion.  The key observation, which
is independent of the direction of $H$ within the hexagonal plane,
is that $T_N(H)$ is almost constant for $\mu_0H\lesssim2$ T and
increases linearly for $\mu_0H\gtrsim2$ T. The derivatives of
isothermal magnetization and magnetostriction (Fig.
\ref{ehrenfest} (d)) both show broad peaks, without any measurable
hysteresis, at $\mu_0H^*\simeq1.5$ T. As seen in Fig.
\ref{phase_diag}, $\mu_0H^*$ is nearly temperature independent,
varying between 1.5 T at 5 K and 2 T near $T_N$. We interpret
$H^*$ as a spin-flop-like crossover field, above which the spins
are aligned nearly perpendicular to the field (Fig. \ref{th_exp}
(e)). Contrary to a first-order spin-flop transition, which occurs
in a single-easy-axis system when $H$ is applied parallel to the
spins, a crossover is obtained because of the presence, at zero
field, of three kinds of domains, in which the spins are
orientated along equivalent hexagonal directions \cite{rogado02}.
In addition to the collective rotation of the spins induced by the
field, a domain alignment, by domain wall motion, probably plays a
crucial role in the alignment of the spins. This crossover may
additionally be broadened because of a tiny domain wall energy due
to the quasi 2D nature of the magnetic exchange \cite{dejongh90}.

For $\mathbf{H}\perp\mathbf{c}$, the alignment of the spins, which
is controlled by the minimization of
$|\mathbf{H}\cdot\mathbf{S}|$, can be represented by an effective
field-induced Ising in-plane anisotropy $D^{eff}_{IP}(\mathbf{H})$
\cite{notespinalignment}, in which the easy-axis is perpendicular
to ($\mathbf{H},\mathbf{c}$). In the high-field AF' state (Fig.
\ref{phase_diag}), i.e. for $\mu_0H\gtrsim2$T,
$D^{eff}_{IP}(\mathbf{H})$ dominates over the intrinsic $D_{IP}$
term and the spins are almost perpendicular to $\mathbf{H}$. The
linear increase of $T_N(H)$ observed in this regime is related to
a reduction of the spin fluctuations along the direction of
$\mathbf{H}$, due to the field-induced anisotropy
\cite{notevillain}. The high-field line extrapolates to $H=0$ at
$T_{N,0}=46.6\pm 0.1$ K, which is smaller than $T_N(H=0)$ by
$\Delta=0.8\pm0.2$ K. Theoretical support is needed to
quantitatively relate $T_{N,0}$ and the slope of $T_N(H)$ to the
characteristic energy scales of the problem. Nevertheless, we
speculate that $T_{N,0}$ is the ordering temperature of the system
in the limit of no in-plane anisotropy, i.e. with
$D_{IP}\rightarrow0$ and $H^*\rightarrow0$, so that $T_{N,0}$ is
only controlled by $J$, $J'$, and $D_{XY}$. In this picture, the
hexagonal Ising-like in-plane anisotropy $D_{IP}$ stabilizes the
long-range magnetic ordering, shifting $T_{N,0}$ upwards to $T_N$
by $\Delta$. Since the 3D character of the long-range ordering was
shown by neutron scattering \cite{rogado02}, we interpret our
phase diagram as resulting from a field-induced crossover from a
3D Ising-like long-range ordering controlled by the hexagonal
in-plane anisotropy $D_{IP}$, to a 3D Ising long-range ordering
controlled by an effective field-induced in-plane anisotropy
$D^{eff}_{IP}(\mathbf{H})$. As long as there is a distribution of
domains, i.e. for $H\lesssim H^*$, $T_N(H)$ is independent of $H$
and is not controlled by $D^{eff}_{IP}(\mathbf{H})$. We speculate
that our phase diagram - and its interpretation - may be generic
to quasi 2D systems with more than one easy axes and with $J\gg
D_{XY}\gg D_{IP}, J'$ \cite{notesuh}.

In the following, we use our data, together with thermodynamic
relationships, to determine the uniaxial pressure dependences of
$T_N$, $J$, and $H^*$. The Ehrenfest relationship $\partial
T_N/\partial p_i=\Delta \alpha_i VT_N/\Delta C_p$, where $\Delta
\alpha_i$ and $\Delta C_p$ are the expansivity and specific heat
jumps at the transition, gives the uniaxial pressure dependences
of $T_N$. Similarly, the uniaxial pressure dependences of $J$ can
be obtained by substituting $T_{max}\sim J$ for $T_N$ in an
Ehrenfest-type relation and considering the total magnetic signal
for $\Delta \alpha$ and $\Delta C_p$. Appropriate scalings of
specific heat and thermal expansion (at $H=0$) are shown in Fig.
\ref{ehrenfest} (a-c) \cite{noteehrenfest}, and the resulting
pressure dependences are listed in Table \ref{table}. The
hierarchy $(1/J)\partial J/\partial p\ll(1/T_N)\partial
T_{N}/\partial p$ indicates that $\partial T_{N}/\partial p$ is
not controlled by $J$. We also note the field-induced sign change
of $\partial T_{N}/\partial p_a$ when $\mathbf{H}\parallel
\mathbf{a}$. In contrast, $\partial T_{N}/\partial p_a$ and
$\partial T_{N}/\partial p_c$ remain positive and negative, when
$\mathbf{H}\perp(\mathbf{p},\mathbf{c})$ and $\mathbf{H}\parallel
\mathbf{a}$, respectively, over the whole investigated $H$ range.
In Fig. \ref{ehrenfest} (d), the derivative of magnetization
$\partial M/\partial (\mu_0H)$, and the magnetostriction
coefficient $\lambda_a^{||}=(1/L_a)\partial L_a/\partial(\mu_0H)$
measured in the longitudinal configuration, are shown at 5 K as a
function of $H$. The data were scaled at the peak values to apply
the generalized Ehrenfest relation $\partial H^*/\partial
p_i=V\Delta\lambda_i/\Delta(\partial M/\partial(\mu_0H))$. The
obtained $p$ dependences of $H^{*}$ (Table \ref{table}) are much
larger than the ones of $T_N$, which in turn are larger than those
of $J$, and they are the largest for uniaxial pressures applied
within the hexagonal plane. Since $H^*$ depends strongly on the
zero field in-plane anisotropy, this suggests that the pressure
dependence of the in-plane anisotropy controls the pressure
dependence of $H^*$. Further, in-plane pressure effects on $H^*$
have opposite signs parallel and perpendicular to the applied
field. In the following discussion, we provide a simple
explanation of the above correlations. We note that the $c$-axis
pressure dependence of $H^*$ is much smaller than the in-plane
pressure dependences of $H^*$ and may result from them via elastic
coupling of the axes.

In the high-field phase (AF' in Fig. \ref{phase_diag}), the spins
are aligned such that their direction is perpendicular to
$\mathbf{H}$. Associated with this spin-flop-like crossover is a
macroscopic distortion of the hexagonal plane, which develops
below $T_N$ and can be extracted from our dilatometry data (the
integration of $\alpha(T)$ and $\lambda(T)$, in Fig. \ref{th_exp}
(a-b)) and \ref{ehrenfest} (d), leads to the length changes). This
in-plane distortion is such that the crystal contracts along the
spin direction and expands perpendicularly to the spins (Fig.
\ref{th_exp} (e)). Since presumably the magnetic domains are
already distorted at low fields \cite{notedistortion}, the
application of uniaxial pressure $\mathbf{p}$ in the easy-plane
will tend to align the spins such that the contracted direction is
parallel to $\mathbf{p}$. Hence, the favored spin direction will
be parallel to $\mathbf{p}$. This effect is analogous to the
magnetic field effect and can be described by adding an effective
pressure-induced in-plane Ising anisotropy term
$D^{eff}_{IP}(\mathbf{p})$ to the Hamiltonian, the corresponding
easy axis being parallel to $\mathbf{p}$.

\begin{table}[t]
\caption{Normalized uniaxial pressure-dependences of $J$, $T_N$,
and $H^*$ obtained from our data using Ehrenfest relations.}
\begin{ruledtabular}
\begin{tabular}{llllc}
&$\mathbf{p}\parallel \mathbf{a}$&$\mathbf{p}\parallel \mathbf{a}$&$\mathbf{p}\parallel \mathbf{c}$\\
(kbar$^{-1}$)&$\mathbf{H}\parallel \mathbf{a}$&$\mathbf{H}\perp(\mathbf{p},\mathbf{c})$&$\mathbf{H}\parallel \mathbf{a}$\\

\hline $(1/J)\partial J/\partial p$\;\;\;\;\;\;\;\;\;\;(0 T)&1.5*10$^{-3}$&1.5*10$^{-3}$&-\\

\hline $(1/T_N)\partial T_{N}/\partial p$\;\;\;\;\;(0 T)&1*10$^{-2}$&1*10$^{-2}$&-6*10$^{-3}$\\

\;\;\;\;\;\;\;\;\;\;\;\;\;\;\;\;\;\;\;\;\;\;\;\;\;\;\;\;(10 T)&-5*10$^{-3}$&1*10$^{-2}$&-4*10$^{-3}$\\

\hline $(1/H^{*})\partial H^{*}/\partial p$&2.5&-1.4&-2.5*10$^{-1}$\\
\end{tabular}
\end{ruledtabular}
\label{table}
\end{table}

The $H$ and $T$ evolutions of this in-plane distortion ultimately
govern the in-plane uniaxial $p$ dependences of $H^*$ and $T_N$.
$\mathbf{p}$ applied along $\mathbf{a}^\perp$, i.e. perpendicular
to $(\mathbf{H},\mathbf{c})$, will favor the alignment of the
spins perpendicular to $\mathbf{H}$, and thus will reduce $H^*$.
Using $(1/H^*)\partial H^{*}/\partial p=-1.4$ kbar$^{-1}$, we
estimate that 0.5 kbar along $\mathbf{a}^\perp$ is enough to
reduce $H^*$ to zero and to align the spins by stress alone. On
the other hand, $(1/H^*)\partial H^{*}/\partial p$ is positive for
$\mathbf{p}$ applied along $\mathbf{a}^\parallel$, i.e. parallel
to $\mathbf{H}$, which means that larger fields will be needed to
align the spins. Magnetic ordering will be favored if both the
field- and the pressure-induced anisotropies
$D^{eff}_{IP}(\mathbf{H})$ and $D^{eff}_{IP}(\mathbf{p})$ act
cooperatively to align the spins along the same axis
\cite{cooperativeeffect}. This explains the positive value of
$\partial T_{N}/\partial p_a^\perp$ at 10 T. In contrast,
$D^{eff}_{IP}(\mathbf{p})$ acts against $D^{eff}_{IP}(\mathbf{H})$
when $\mathbf{p}$ is applied along $\mathbf{a}^\parallel$, i.e.
parallel to $\mathbf{H}$, which leads to $\partial T_{N}/\partial
p_a^\parallel<0$ at 10 T.

In conclusion, the ($T,H$) phase diagram of
BaNi$_{2}$V$_{2}$O$_{8}$ was studied with
$\mathbf{H}\perp\mathbf{c}$. The obtained field- and uniaxial
pressure-dependences of $T_N$ were interpreted as the consequence
of effective field- and pressure-induced in-plane anisotropies.
Our data clearly demonstrate the importance of the in-plane
anisotropy for establishing 3D long-range order in
BaNi$_{2}$V$_{2}$O$_{8}$. A search for BKT behavior should be
restricted to a temperature region above $T_N$, where in-plane
anisotropy no longer influences the growth of the 2D-XY
correlations. In the limit of $D_{IP}=0$, we extrapolated a N\'eel
temperature $T_{N,0}=46.6$ K, which provides an upper estimate of
the BKT temperature $T_{BKT}$ \cite{noteBKT}. Since the Ising-like
anisotropy $D_{IP}$ leads to an increase of $T_N$ by almost 1 K,
2D-XY BKT behavior may be valid for temperatures at least several
degrees higher than $T_N$. These results are expected to be useful
for future studies of BKT behavior.

We acknowledge useful discussions with L.P. Regnault, C. Boullier,
D. Reznik, T. Roscilde, J. Villain, S. Bayrakci, B. Keimer, and R.
Eder. This work was supported by the Helmholtz-Gemeinschaft
through the Virtual Institute of Research on Quantum Phase
Transitions and Project VH-NG-016.


\begin{thebibliography}{20}


\bibitem{dejongh90} {\it Magnetic properties of layered
transition metal compounds}, edited by L.J. DeJongh (Kluwer
Academic Publishers, Dordrecht/Boston/London, 1990).

\bibitem{berezinskii71kosterlitz73}
V.L. Berezinskii, Sov. Phys. JETP {\bf 32}, 493 (1971); J.M.
Kosterlitz and D.J. Thouless, J. Phys. C {\bf 6}, 1181 (1973).

\bibitem{minnhagen87} P. Minnhagen, Rev. Mod. Phys. {\bf59}, 1001
(1987).

\bibitem{cuccoli03}
A. Cuccoli et al., Phys. Rev. B {\bf67}, 104414 (2003).

\bibitem{rogado02} N. Rogado et al., Phys. Rev. B {\bf65}, 144443
(2002).

\bibitem{heinrich03}
M. Heinrich et al., Phys. Rev. Lett. {\bf91}, 137601 (2003).

\bibitem{reznik06} Two spin wave gaps $\Delta_{XY}\simeq3$ and $\Delta_{IP}\simeq0.2$ meV,
related to the XY and in-plane anisotropies, respectively, were
recently measured by neutron scattering [D. Reznik, S. Bayrakci,
J. Lynn, and B. Keimer, {\it private communication}]; using
$J\simeq10$ meV and $\Delta_i\propto\sqrt{D_{i}J}$, we estimate
$D_{XY}\simeq1$ and $D_{IP}\simeq 4.10^{-3}$ meV (to our
knowledge, no experiment was able to estimate $J'$).

\bibitem{knafo07} In [W. Knafo et al., J. Magn. Magn. Mater.
{\bf310}, 1248 (2007)], a slight increase of $T_N$ for
$\mathbf{H}\parallel\mathbf{c}$ was associated to a field-induced
XY anisotropy.

\bibitem{lashley03} The procedure proposed in
[J.C. Lashley et al., Cryogenics {\bf43}, 369 (2003)] was used to
enhance resolution.

\bibitem{notephonons}
The phonon contribution is assumed to be similar in both
BaNi$_{2}$V$_{2}$O$_{8}$ and BaNi$_{0.8}$Mg$_{1.2}$V$_{2}$O$_{8}$,
while the magnetic contribution of
BaNi$_{0.8}$Mg$_{1.2}$V$_{2}$O$_{8}$ has a weight only at very low
temperatures.

\bibitem{noterogado}
We recall that in Ref. \cite{rogado02}, a "small feature" was
observed at $T_N\simeq48$ K in the specific heat, the associated
entropy being estimated by $\Delta S_N/\Delta S_{mag} \simeq
8*10^{-3}$.

\bibitem{bloembergen77} P. Bloembergen, Physica  {\bf85B}, 51
(1977).

\bibitem{notespinalignment} Here, the spins align $\perp\mathbf{H}$; ultimately, for $\mu_0H\gg10$ T,
the antiferromagnetic phase will be replaced by a polarized state
with the spins $\parallel\mathbf{H}$. An introduction about
effective field-induced anisotropy can be found in Ref.
\cite{dejongh90}.

\bibitem{notevillain} This picture has been proposed in [J. Villain and
J. M. Loveluck, J. Phys. (Paris) {\bf38}, L77 (1977)] for quasi 1D
magnetic systems in a magnetic field.

\bibitem{notesuh}
Such phase diagrams were already obtained in [I.W. Sumarlin et
al., Phys. Rev. B {\bf51}, 5824 (1995)] and [B.J. Suh et al.,
Phys. Rev. Lett. {\bf75}, 2212 (1995)] on the quasi 2D tetragonal
magnetic systems Pr$_2$CuO$_4$ and Sr$_2$CuO$_2$Cl$_2$,
respectively. While Sumarlin et al. did not discuss the increase
of $T_N(H)$, Suh et al. introduced a picture with a field-induced
crossover from XY- to Ising-driven N\'eel ordering, which is quite
different from our interpretation of the phase diagram. In [Y.
Shapira et al., Phys. Rev. B {\bf21}, 1271 (1980)], a similar
phase diagram was obtained for the 3D hexagonal CsMnF$_3$, being
interpreted as a XY to Ising crossover (as in Suh et al.). An
increase of $T_N(H)$ was also reported for quasi 2D systems with a
single easy axis \cite{dejongh90}, although their phase diagram is
somewhat different than ours (first order spin-flop transition
instead of a spin alignment crossover).


\bibitem{noteehrenfest}
The magnetic contribution to the thermal expansion of Fig.
\ref{ehrenfest} (a) is estimated using
$\alpha_{mag}=\alpha-\alpha_{ph}$, where $\alpha_{ph}=C_{ph}*s$ is
the phononic contribution, $s$ being a constant refined so that
$\alpha_{mag}(T)$ and $C_{mag}(T)$ have the same shape. The peaks
at $T_N$ plotted in Fig. \ref{ehrenfest} (b) and (c) were
extracted using appropriate backgrounds.

\bibitem{notedistortion}
A distortion relative to the spin direction characterizes
certainly each magnetic domain at $H=0$, but is not observed
macroscopically because of averaging over all domains. For
$\mathbf{H}\perp\mathbf{c}$, the formation of a single domain due
to spin alignment is believed to induce the macroscopic distortion
observed experimentally.


\bibitem{cooperativeeffect}
Here both $\mathbf{p}\parallel \mathbf{a}$ and
$\mathbf{H}\perp(\mathbf{a},\mathbf{c})$ favor an alignment of the
spins parallel to $\mathbf{a}$.

\bibitem{noteBKT} This is compatible with $T_{BKT}=43.3$ K estimated from a fit of the
ESR linewidth in Ref. \cite{heinrich03} (where $T_N\simeq50$ K was
also obtained, instead of 47.4 K here).

\end{thebibliography}
\end{document}